\documentclass[3p,times,procedia]{elsarticle}
\usepackage{nupha_ecrc}

\volume{00}

\firstpage{1}

\journalname{Nuclear Physics A}

\runauth{}

\jid{nupha}

\jnltitlelogo{Nuclear Physics A}

\usepackage{amssymb}

\usepackage[figuresright]{rotating}

\begin{document}

\begin{frontmatter}



\dochead{XXVIIIth International Conference on Ultrarelativistic Nucleus-Nucleus Collisions\\ (Quark Matter 2019)}

\title{QCD equation of state at finite densities for nuclear collisions}


\author[label1,e1]{Akihiko Monnai}
\author[label2]{Bj\"orn Schenke}
\author[label3,label4]{Chun Shen}

\address[label1]{KEK Theory Center, Institute of Particle and Nuclear Studies, \\
High Energy Accelerator Research Organization, Tsukuba, Ibaraki 305-0801, Japan}
\address[label2]{Physics Department, Brookhaven National Laboratory, Upton, NY 11973, USA}
\address[label3]{Department of Physics and Astronomy, Wayne State University, Detroit, Michigan, USA}
\address[label4]{RIKEN BNL Research Center, Brookhaven National Laboratory, Upton, NY 11973, USA}

\address{}

\begin{abstract}
We construct the QCD equation of state at finite chemical potentials including net baryon, electric charge, and strangeness based on the results of lattice QCD simulations and the hadron resonance gas model. The situation of strangeness neutrality and a fixed charge-to-baryon ratio, which resembles that of heavy nuclei, is considered for the application to relativistic heavy-ion collisions. This increases the values of baryon chemical potential compared to the case of vanishing strangeness and electric charge chemical potentials, modifying the fireball trajectory in the phase diagram. We perform  viscous hydrodynamic simulations and demonstrate the importance of multiple chemical potentials for identified particle production in heavy-ion collisions at the RHIC and SPS beam energy scan energies.
\end{abstract}


\end{frontmatter}


\section{Introduction}
\label{}
The properties of the quark-gluon plasma (QGP) 
have been explored extensively 
at the BNL Relativistic Heavy Ion Collider (RHIC) and CERN Large Hadron Collider (LHC). One of the discoveries in the collider experiments has been the nearly perfect fluidity of the produced quark matter. 
This implies that the thermodynamic properties of QCD is relevant in understanding 
relativistic nuclear collisions. 

The equation of state is a fundamental relation among thermodynamic variables. First principle calculations based on the lattice QCD method have been successful in determining the QCD equation of state at vanishing chemical potentials. Such estimations indicate that the quark-hadron transition is a crossover in the zero baryon density limit. On the other hand, the method suffers from the sign problem at finite chemical potentials. 
The Beam Energy Scan (BES) program is being performed at RHIC and CERN Super Proton  Synchrotron (SPS) and also planned at various other collider facilities, seeking experimental insight into the phase structure of QCD, including the conjectured critical point \cite{Asakawa:1989bq}. 

We construct the equation of state with net baryon ($B$), electric charge ($Q$) and strangeness ($S$) using the pressure and susceptibilities of lattice QCD simulations and the hadron resonance gas model \cite{Monnai:2019hkn, Noronha-Hostler:2019ayj}.
We then use it in hydrodynamic analyses to demonstrate its effects on the description of particle production.

\section{Construction of the equation of state}

In our model \textsc{neos}, the equation of state is constructed by connecting the pressure of the Taylor expansion method in lattice QCD to that of a hadron resonance gas on the lower temperature side \cite{Monnai:2019hkn,Monnai:2015sca}. The procedure is motivated by the facts that (i) the Taylor expansion is not reliable when the fugacity is large, (ii) the pressure and susceptibilities of lattice QCD and hadron resonance gas show reasonable agreement around and below the crossover temperature and (iii) the hadron resonance gas equation of state is encoded in the Cooper-Frye prescription of fluid particliazation, which is used at the last stage of hydrodynamic evolution.

The pressure in the QCD system with $u,d$ and $s$ quarks is expressed in the Taylor expansion method as
\begin{eqnarray}
\frac{P}{T^4} = \frac{P_0}{T^4} + \sum_{l,m,n} \frac{\chi^{B,Q,S}_{l,m,n}}{l!m!n!} \bigg( \frac{\mu_B}{T} \bigg)^{l} \bigg( \frac{\mu_Q}{T} \bigg)^{m}  \bigg( \frac{\mu_S}{T} \bigg)^{n}, 
\label{Psus}
\end{eqnarray}
where $P$ is the pressure, $T$ is the temperature, $\chi^{B,Q,S}_{l,m,n}$ are the ($l$+$m$+$n$)-th order susceptibilities and $\mu_{B,Q,S}$ are the chemical potentials for net baryon, electric charge, and strangeness. $P_0$ denotes $P(\mu_{B,Q,S}=0)$. 
We employed the latest results for the pressure and the second- and fourth-order susceptibilities of (2+1)-flavor lattice QCD simulations \cite{Bazavov:2014pvz,Bazavov:2012jq, Ding:2015fca, Bazavov:2017dus}. 

The hadron resonance gas equation of state, on the other hand, is given as
\begin{eqnarray}
P = \pm T \sum_i \int \frac{g_i d^3p}{(2\pi)^3} \ln [1 \pm e^{-(E_i-\mu_i)/T} ],
\label{eq:P_had}
\end{eqnarray}
where $i$ is the index of hadron species, $g_i$ is the degeneracy, $E_i$ is the energy and $\mu_i = B_i \mu_B + Q_i \mu_Q + S_i \mu_S$ is the chemical potential of the hadron. Here $B_i$, $Q_i$ and $S_i$ are the quantum numbers for net baryon, electric charge, and strangeness, respectively.
The upper/lower signs are for fermions/bosons.
All hadron resonances in the particle data group list \cite{Tanabashi:2018oca} with the quark components $u,d,s$ and mass below 2~GeV are considered. 

The hybrid equation of state is constructed by connecting the two equations of state as
\begin{eqnarray}
\frac{P}{T^4} = \frac{1}{2}\bigg[1-\tanh \frac{T-T_c(\mu_B)}{\Delta T_c}\bigg] \frac{P_{\mathrm{had}}}{T^4}
+ \frac{1}{2}\bigg[1+\tanh \frac{T-T_c(\mu_B)}{\Delta T_c} \bigg] \frac{P_{\mathrm{lat}}}{T^4} , \label{eq:econ}
\end{eqnarray}
where $T_c(\mu_B) = 0.16\ \mathrm{GeV} - 0.4\times(0.139\ \mathrm{GeV}^{-1} \mu_B^2 + 0.053\ \mathrm{GeV}^{-3} \mu_B^4)$ based on the chemical freeze-out curve \cite{Cleymans:2005xv} and $\Delta T_c = 0.1 T_c(0)$ are used. The procedure yields the equation of state with a crossover.
We impose the thermodynamic conditions $\partial^2 P/\partial T^2 = \partial s/\partial T > 0$ and $\partial^2 P/\partial \mu_{B,Q,S}^2 = \partial n_{B,Q,S}/\partial \mu_{B,Q,S} > 0$
as constraints to the parameters. 
The Stefan-Boltzmann limit is used to regulate the high-temperature behavior of the pressure. 
We phenomenologically introduce $\chi_6^B$, $\chi_{5,1}^{B,Q}$, and $\chi_{5,1}^{B,S}$, which are most relevant among the sixth-order susceptibilities to $n_B$, $n_Q$, and $n_S$, in the QGP phase to ensure that they are smooth functions of $T$ and $\mu_B$ and that the connection preserves the results of the hadron resonance gas model below $T_c$.

\section{Numerical results}

We consider three conditions (i) $\mu_S$ = $\mu_Q$ = 0 (\textsc{neos} B), (ii) $n_S$ = 0 and $\mu_Q$ = 0 (\textsc{neos} BS) and (iii) $n_S$~=~0 and $n_Q$ = $0.4n_B$ (\textsc{neos} BQS) for the equation of state. 
The first is the conventional case where only a net baryon chemical potential is considered, the second is the case with the strangeness neutrality condition, and the third is the case that also takes into account the charge-to-baryon ratio in heavy nuclei. Note that the colliding nuclei do not have net strangeness and the proton-to-nucleon number ratio is $0.401$ for ${}^{197}_{\ 79}$Au and $0.394$ for ${}^{208}_{\ 82}$Pb. The conditions can be violated in the presence of initial fluctuation or diffusion processes.

\begin{figure}[tb]
\begin{center}
\includegraphics[width=1.95in]{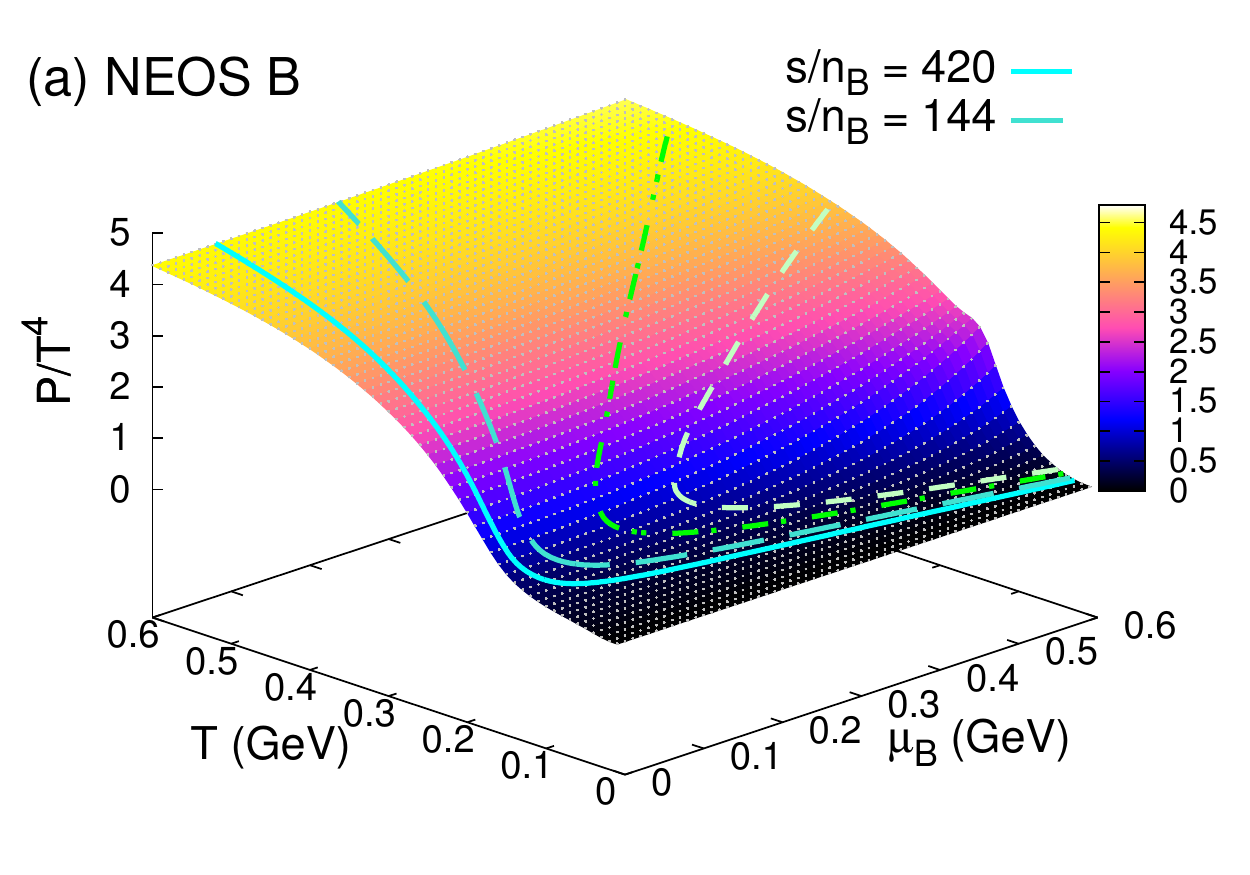}
\includegraphics[width=1.95in]{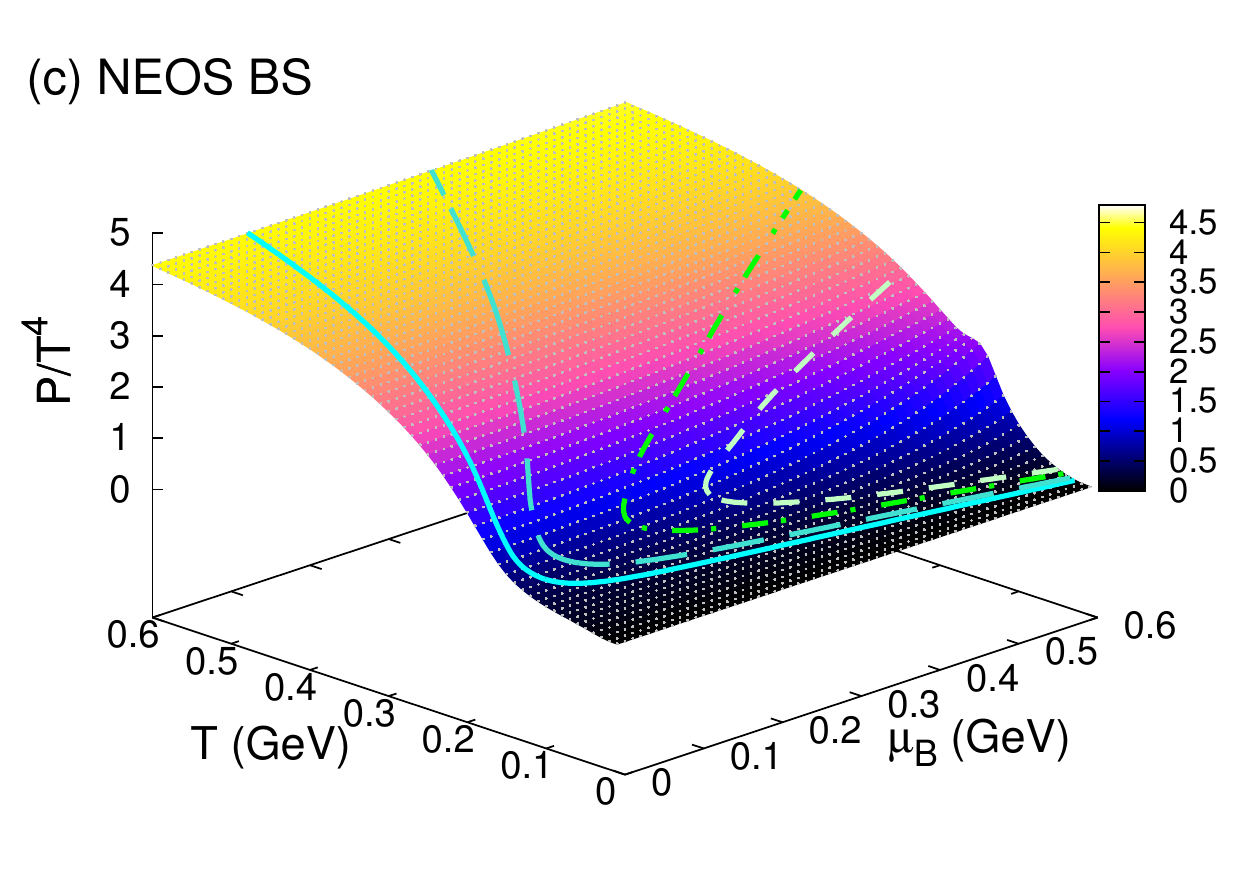}
\includegraphics[width=1.95in]{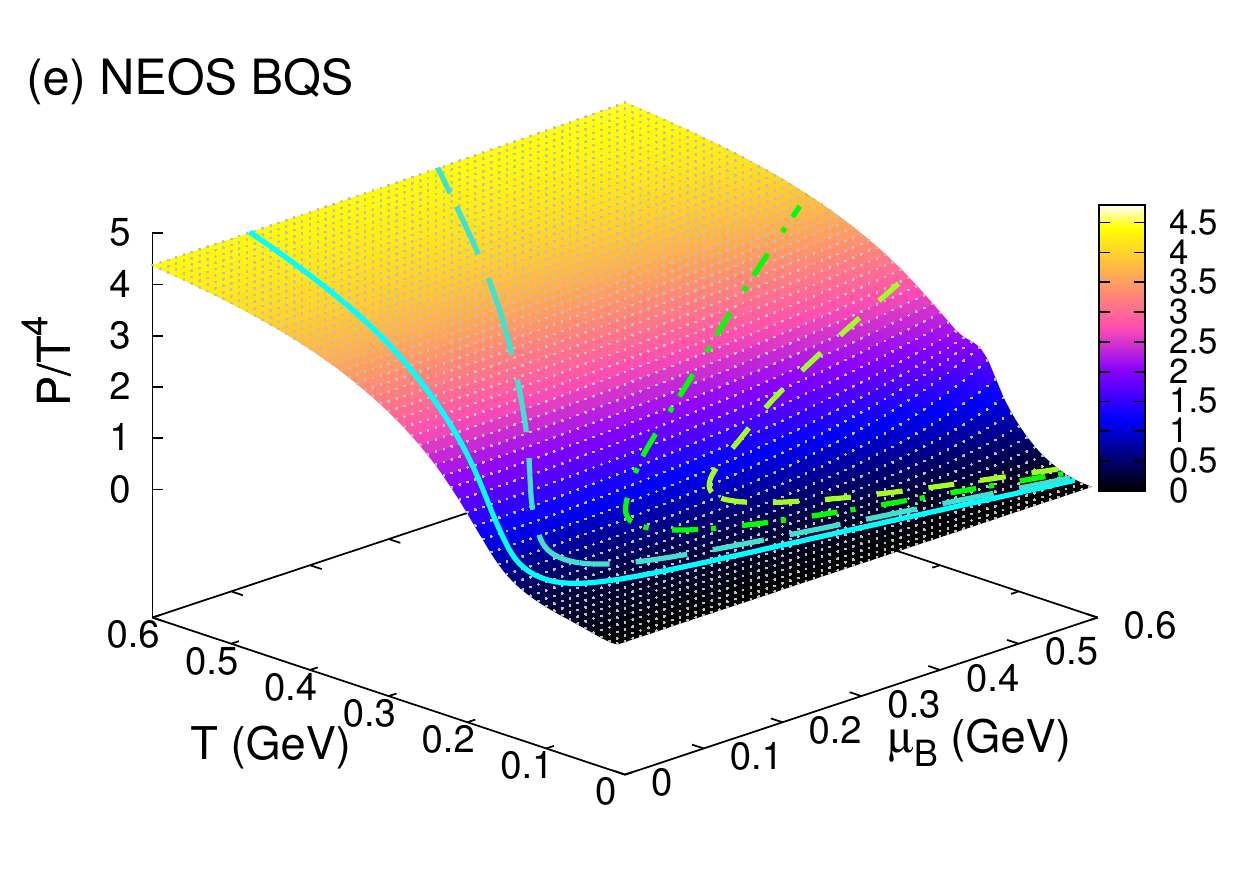}
\includegraphics[width=1.95in]{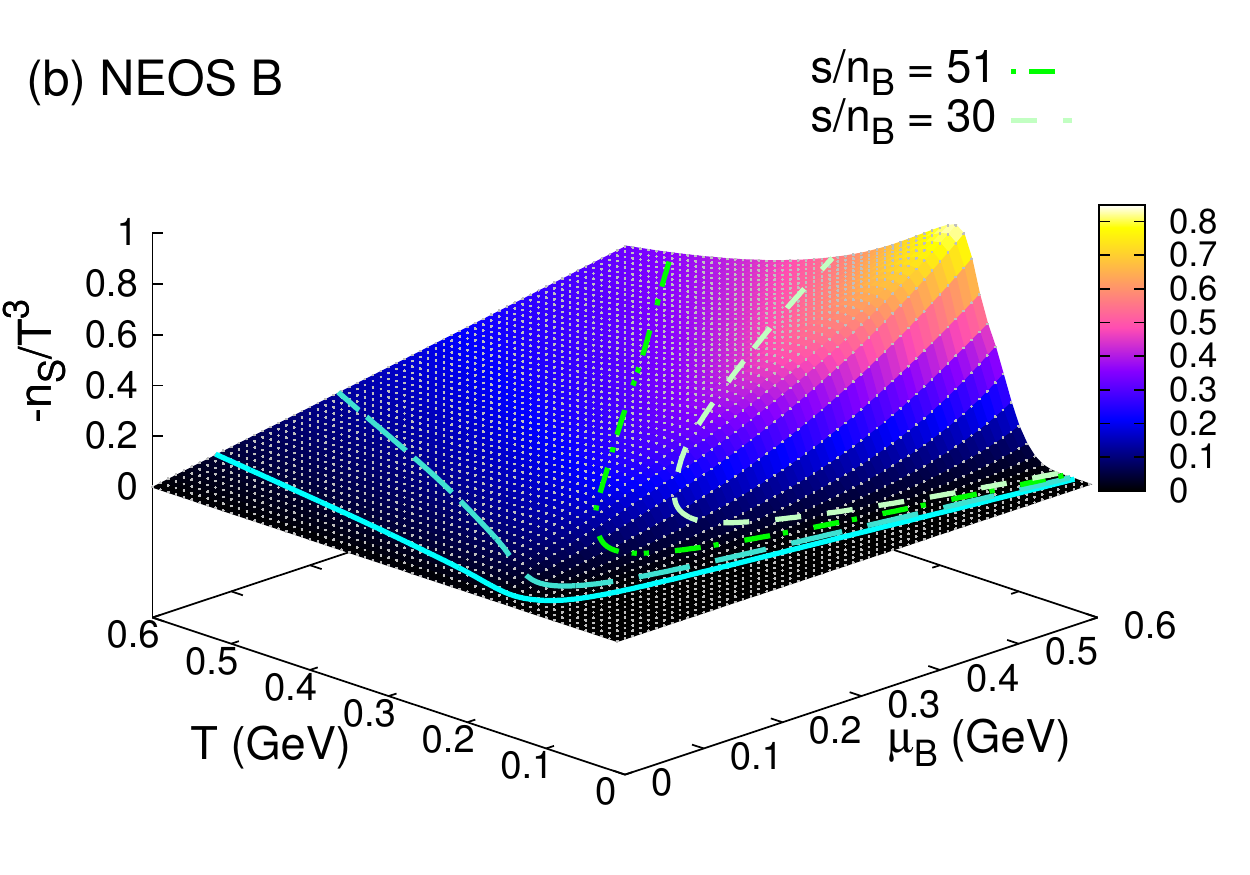}
\includegraphics[width=1.95in]{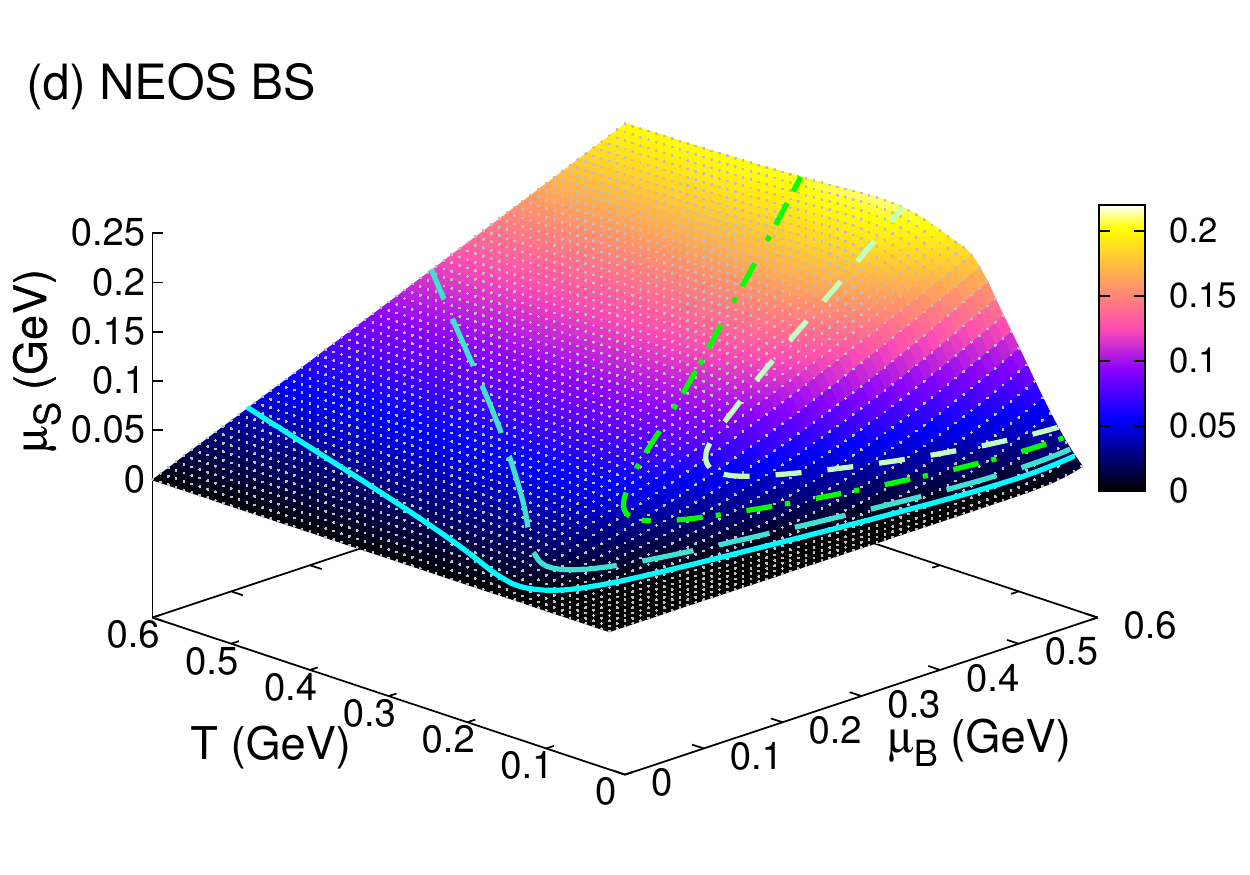}
\includegraphics[width=1.95in]{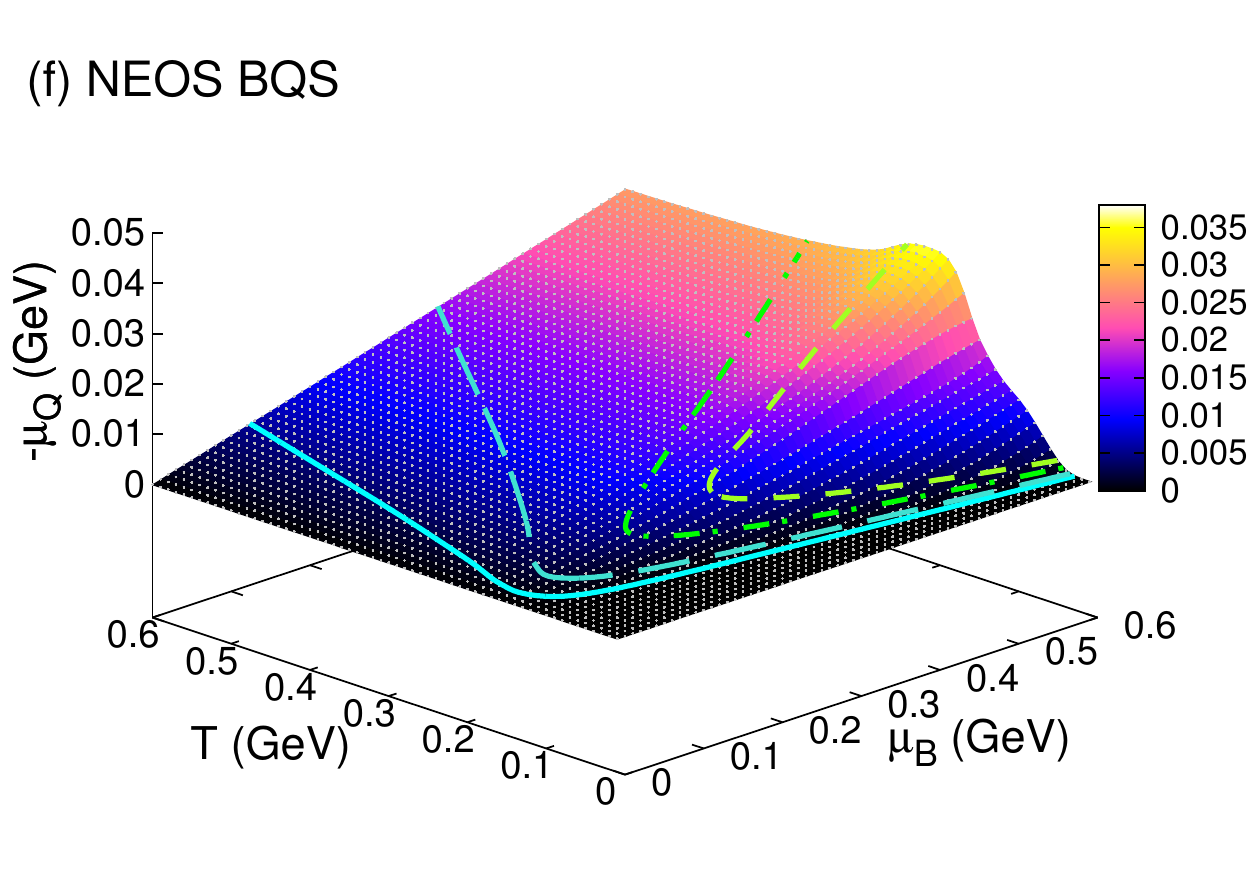}
\end{center}
\caption{(a) $P/T^4$ and (b) -$n_S/T^3$ in \textsc{neos} B, (c) $P/T^4$ and (d) $\mu_S$ in \textsc{neos} BS, and (e) $P/T^4$ and (f) -$\mu_Q$ in \textsc{neos} BQS as functions of $T$ and $\mu_B$. The solid, long-dashed, dash-dotted, and short-dashed lines are the trajectories for different constant $s/n_B$.}
\label{fig:neos}
\end{figure}

The pressure as a function of the temperature and baryon chemical potential for \textsc{neos} B is shown in Fig.~\ref{fig:neos} (a). The pressure is a monotonically increasing function of $T$ and $\mu_B$. The $s/n_B$ ratio is approximately fixed when the entropy and net baryon number are conserved and thus its trajectory shows the typical probe region for each collision energy. $s/n_B$ = 420, 144, 51 and 30 correspond to $\sqrt{s_{NN}}$ = 200, 62.4, 19.6 and 14.5 GeV, respectively \cite{Gunther:2016vcp}. One can see that the region of least reliability where $\mu_B/T >3$ above $T_c$ tends to be avoided. The strangeness neutrality condition is violated by \textsc{neos} B as seen in Fig.~\ref{fig:neos} (b). 

In \textsc{neos} BS where $n_S = 0$ and $\mu_Q = 0$, the pressure is modified at larger $\mu_B$ near $T_c$ as shown in Fig.~\ref{fig:neos} (c). The strangeness chemical potential is finite and positive in this case (Fig.~\ref{fig:neos} (d)). In the parton gas limit, this can be understood with the condition for $s$ quark chemical potential $\mu_s = \mu_B/3 - \mu_S = 0$ when $\mu_Q = 0$. Since only $u$ and $d$ can contribute to the net baryon density in the QGP phase, $\mu_B$ becomes larger for a given $T$, pushing the fireball trajectories of constant $s/n_B$ to larger $\mu_B$. This may affect the estimation of baryon diffusion current which is a response to the spacial gradient in $\mu_B/T$.

The pressure with the strangeness neutrality condition and the realistic charge-to-baryon ratio is presented in Fig.~\ref{fig:neos} (e). Although the change from \textsc{neos} BS to \textsc{neos} BQS is small, it can be important in the interpretation of charged particle production. The electric chemical potential is finite and negative (Fig.~\ref{fig:neos} (f)) reflecting the neutron-rich nature of heavy nuclei, which is interpreted in the parton gas picture as the $d$ quark abundance $\mu_d = \mu_B/3 - \mu_Q/3 > \mu_u = \mu_B/3 + 2\mu_Q/3$. $\mu_Q$ becomes positive for proton-rich nuclei. 

We next use the equation of state in the hydrodynamic model to estimate the effects of multiple conserved charges on identified particle production in heavy-ion collisions. Numerical results are compared with the experimental data for Pb+Pb collisions at $\sqrt{s_{NN}}$ = 17.3 GeV at SPS.
The initial conditions are calculated using the dynamical Glauber Monte Carlo model \cite{Shen:2017bsr}. The hydrodynamic model \textsc{Music} \cite{Schenke:2010rr} is used for the medium evolution with shear viscosity $\eta/s = 0.08$ and switched to the UrQMD model \cite{Bass:1998ca,Bleicher:1999xi} at energy density $e_\mathrm{sw} = 0.26$~GeV/fm$^3$.

\begin{figure}[tb]
\begin{center}
\includegraphics[width=2.4in]{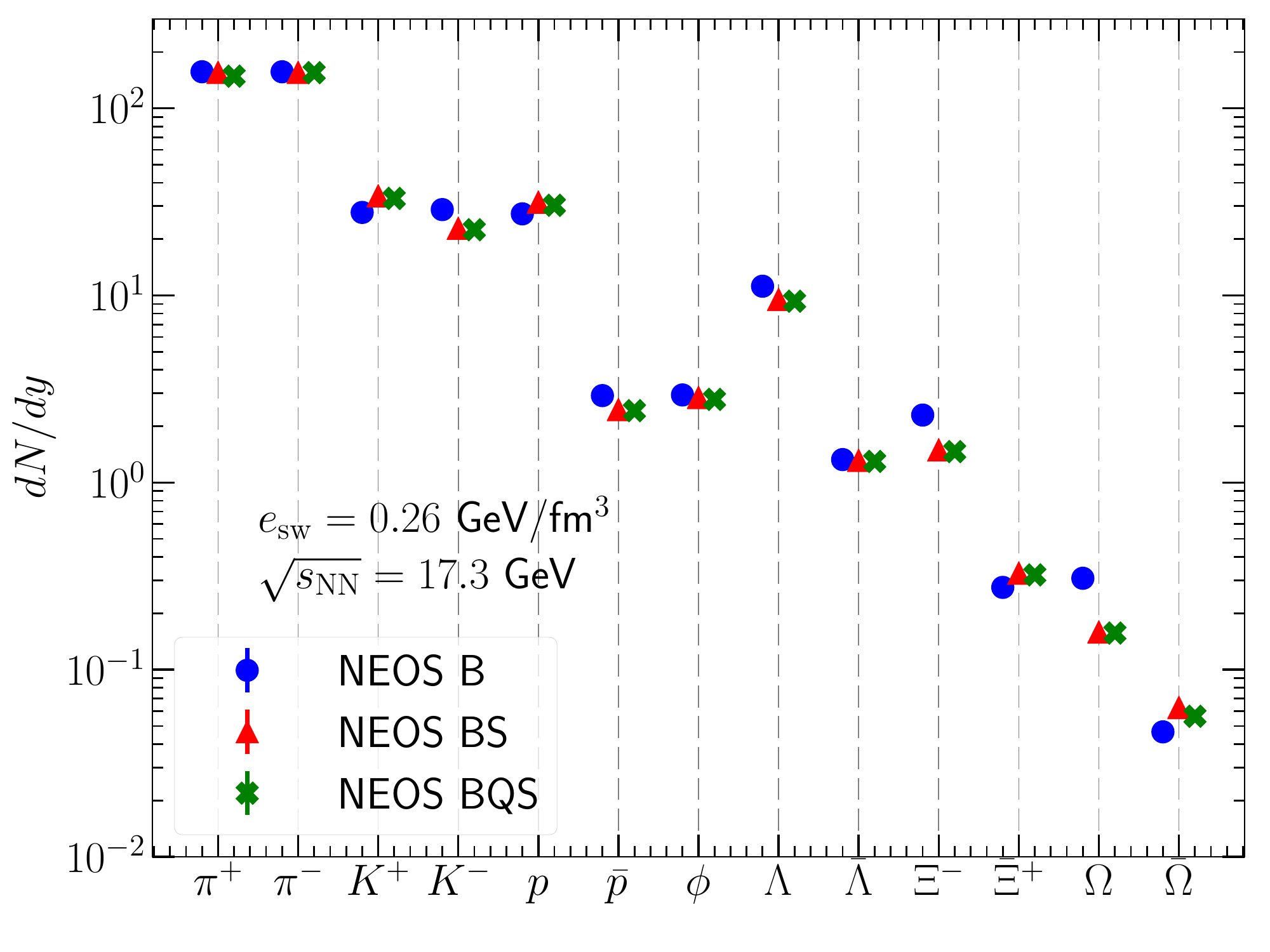}
\includegraphics[width=2.4in]{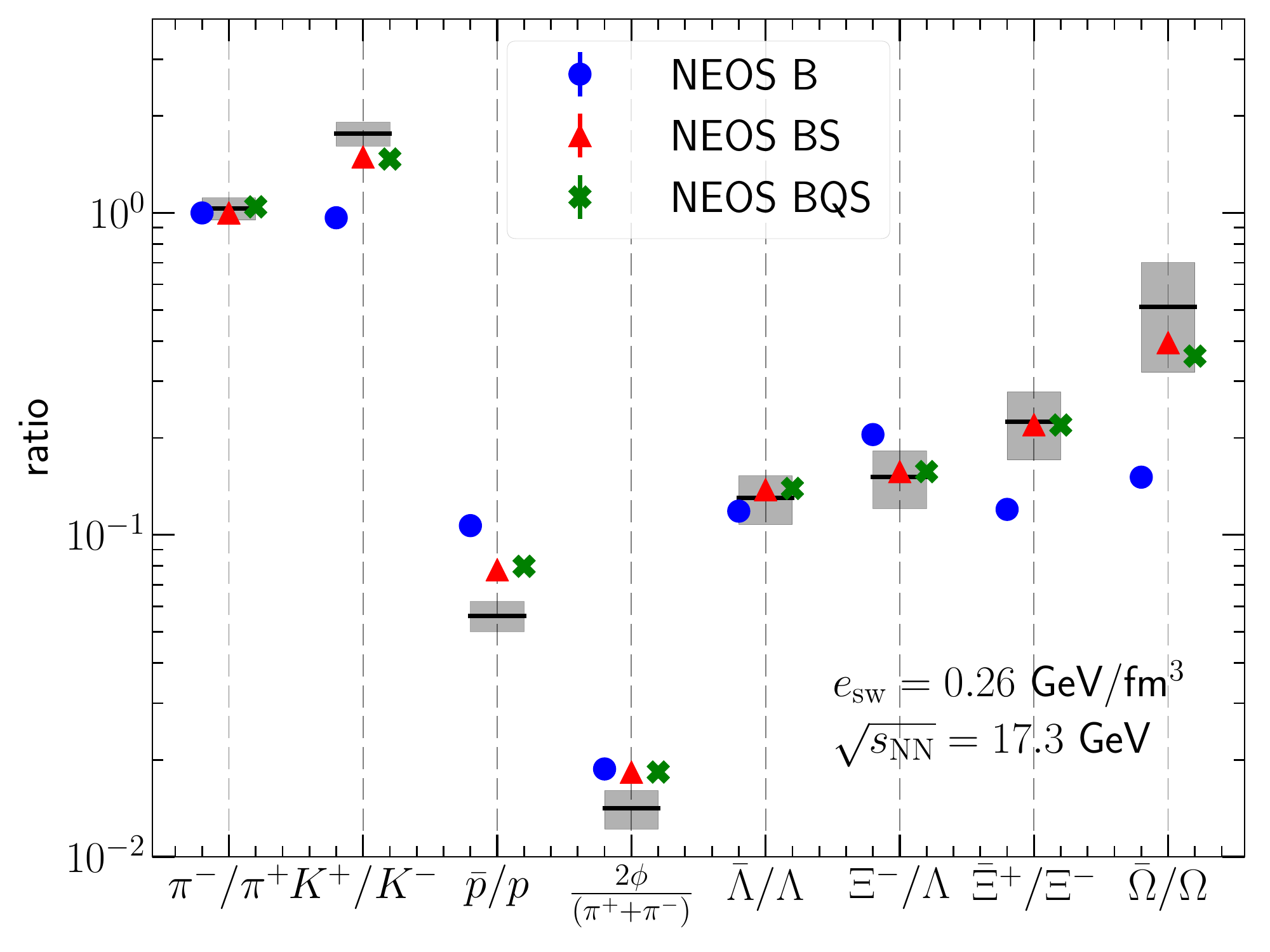}
\end{center}
\caption{(Left) Particle yields and (right) their ratios for the three equations of state compared to the experimental data 
compiled in \cite{NA49data}.}
\label{fig:yield}
\end{figure}

The particle yields and ratios with three different equations of state are shown in Fig.~\ref{fig:yield}. 
One can see that the strangeness neutrality condition visibly improves the quantitative description of the data, especially for the hadrons with strangeness. The $\bar{p}/p$ ratio is also improved owing to the aforementioned enhancement in $\mu_B$. Introduction of the charge-to-baryon ratio has a small effect, but is essential for understanding charged particle ratios, \textit{e.g.}, $\pi^-/\pi^+$, which is larger than unity. Our simulations indicate that $e_\mathrm{sw} = 0.16$-$0.36$~GeV/fm$^3$ is preferred over values outside that range.

\section{Conclusions}

We have constructed the QCD equation of state at finite chemical potentials of net baryon, electric charge, and strangeness by matching the results of the lattice QCD method and hadron resonance gas model. We have demonstrated in the hydrodynamic model analysis that the description of particle production is improved in the presence of multiple chemical potentials. Our results imply that the $T$-$\mu_B$-$\mu_Q$-$\mu_S$ space rather than the $T$-$\mu_B$ plane is explored in the BES experiments, possibly affecting the theoretical interpretations of the critical point search. The charge-to-baryon ratio has to be modified for the collisions of different nuclei, \textit{e.g.} small systems and isobar experiments. The results of \textsc{neos} are publicly available \cite{neos}.

\section*{Acknowledgments}
The authors thank Frithjof Karsch, Swagato Mukherjee, and Sayantan Sharma for useful discussion. AM is supported by JSPS KAKENHI Grant Number JP19K14722. BPS is supported under DOE Contract No. DE-SC0012704. CS is supported under DOE Contract No. DE-SC0013460. This research used resources of the National Energy Research Scientific Computing Center, which is supported by the Office of Science of the U.S. Department of Energy under Contract No. DE-AC02-05CH11231. This work is supported in part by the U.S. Department of Energy, Office of Science, Office of Nuclear Physics, within the framework of the Beam Energy Scan Theory (BEST) Topical Collaboration.

\bibliographystyle{elsarticle-num}
\bibliography{neos_qm19}

\begin{thebibliography}{10}
\expandafter\ifx\csname url\endcsname\relax
  \def\url#1{\texttt{#1}}\fi
\expandafter\ifx\csname urlprefix\endcsname\relax\def\urlprefix{URL }\fi
\expandafter\ifx\csname href\endcsname\relax
  \def\href#1#2{#2} \def\path#1{#1}\fi

\bibitem{Asakawa:1989bq}
M.~Asakawa, K.~Yazaki, {Chiral Restoration at Finite Density and Temperature},
  Nucl. Phys. A504 (1989) 668--684.

\bibitem{Monnai:2019hkn}
A.~Monnai, B.~Schenke, C.~Shen, {Equation of state at finite densities for QCD
  matter in nuclear collisions}, Phys. Rev. C100~(2) (2019) 024907.

\bibitem{Noronha-Hostler:2019ayj}
J.~Noronha-Hostler, P.~Parotto, C.~Ratti, J.~M. Stafford, {Lattice-based
  equation of state at finite baryon number, electric charge and strangeness
  chemical potentials}, Phys. Rev. C100~(6) (2019) 064910.

\bibitem{Monnai:2015sca}
A.~Monnai, B.~Schenke, {Pseudorapidity correlations in heavy ion collisions
  from viscous fluid dynamics}, Phys. Lett. B752 (2016) 317--321.

\bibitem{Bazavov:2014pvz}
A.~Bazavov, et~al., {Equation of state in (2+1)-flavor QCD}, Phys. Rev. D90
  (2014) 094503.

\bibitem{Bazavov:2012jq}
A.~Bazavov, et~al., {Fluctuations and Correlations of net baryon number,
  electric charge, and strangeness: A comparison of lattice QCD results with
  the hadron resonance gas model}, Phys. Rev. D86 (2012) 034509.

\bibitem{Ding:2015fca}
H.~T. Ding, S.~Mukherjee, H.~Ohno, P.~Petreczky, H.~P. Schadler, {Diagonal and
  off-diagonal quark number susceptibilities at high temperatures}, Phys. Rev.
  D92~(7) (2015) 074043.

\bibitem{Bazavov:2017dus}
A.~Bazavov, et~al., {The QCD Equation of State to $\mathcal{O}(\mu_B^6)$ from
  Lattice QCD}, Phys. Rev. D95~(5) (2017) 054504.

\bibitem{Tanabashi:2018oca}
M.~Tanabashi, et~al., {Review of Particle Physics}, Phys. Rev. D98~(3) (2018)
  030001.

\bibitem{Cleymans:2005xv}
J.~Cleymans, H.~Oeschler, K.~Redlich, S.~Wheaton, {Comparison of chemical
  freeze-out criteria in heavy-ion collisions}, Phys. Rev. C73 (2006) 034905.

\bibitem{Gunther:2016vcp}
J.~N. Guenther, et~al., {The QCD equation of state at finite density from
  analytical continuation}, Nucl. Phys. A967 (2017) 720--723.

\bibitem{Shen:2017bsr}
C.~Shen, B.~Schenke, {Dynamical initial state model for relativistic heavy-ion
  collisions}, Phys. Rev. C97~(2) (2018) 024907.

\bibitem{Schenke:2010rr}
B.~Schenke, S.~Jeon, C.~Gale, {Elliptic and triangular flow in event-by-event
  (3+1)D viscous hydrodynamics}, Phys. Rev. Lett. 106 (2011) 042301.

\bibitem{Bass:1998ca}
S.~A. Bass, et~al., {Microscopic models for ultrarelativistic heavy ion
  collisions}, Prog. Part. Nucl. Phys. 41 (1998) 255--369.

\bibitem{Bleicher:1999xi}
M.~Bleicher, et~al., {Relativistic hadron hadron collisions in the
  ultrarelativistic quantum molecular dynamics model}, J. Phys. G25 (1999)
  1859--1896.

\bibitem{NA49data}
https://edms.cern.ch/document/1075059.

\bibitem{neos}
https://sites.google.com/view/qcdneos/.

\end{thebibliography}

\end{document}